\documentclass[
    ,final            
  ]
  {aipproc}

\layoutstyle{6x9}

\begin{document}

\title{Recent status of polarized parton distributions}

\author{M. Hirai \\ \vspace{1mm} (Asymmetry Analysis Collaboration )
\footnote{A fortran program of the AAC PDF library is available from 
http://spin.riken.bnl.gov/aac/.}
}
{
address={Radiation Laboratory, 
RIKEN (The institute of Physics and Chemical Research), \\
 Wako, Saitama 351-0198, Japan}
}

\begin{abstract}
We study an influence of precise data on uncertainty 
of polarized parton distribution functions. 
This analysis includes the SLAC-E155 proton target data 
which are precise measurements. 
Polarized PDF uncertainties are estimated by using the Hessian matrix. 
We examine correlation effect between 
the antiquark and gluon uncertainties. 
It suggests that reducing the gluon uncertainty is needed 
to determine the polarized antiquark distribution clearly.
\end{abstract}

\maketitle

\section{Introduction}
Polarized parton distribution functions (polarized PDF's) 
have so far been optimized 
from polarized deep inelastic scattering (polarized DIS) world data 
\cite{AAC, polpdf}. 
We could obtain only a slight piece of information 
about polarized antiquark and gluon distributions. 
At this stage, the antiquark SU(3)$_f$ flavor symmetry is assumed 
in most of the polarized PDF analyses. 
The SU(3)$_f$ symmetry breaking is already known 
as the Gottfried sum rule violation in the unpolarized case. 
In principle, the polarized PDF analysis should take account of the symmetry breaking. 
However, we must not only determine a shape 
but also a sign of each polarized antiquark distribution. 
It needs more precise data to improve the current status.
Semi-inclusive DIS experiments \cite{SDIS} are also expected 
to separate antiquark flavor distributions. 
However, the separated distributions may not be credible 
due to ambiguity of the fragmentation functions. 
Then, antiquark flavor distributions cannot be decomposed clearly. 
The current knowledge of the polarized gluon distribution is still poor.
The polarized gluon distribution is suggested as the positive distribution;
however, there is large difference between various parameterization results.

We would like to know ambiguity of polarized PDF's quantitatively.
PDF uncertainty plays an important role in illustrating the ambiguity.
Furthermore, it is important to show the phenomenological uncertainty 
of predicted physical quantities 
(e.g., scattering cross-sections and spin asymmetries) 
with parameterized PDF's and their uncertainties in our work. 
A purpose of this analysis is to clarify the current knowledge 
about the polarized PDF's from the polarized DIS world data 
by using their PDF uncertainty.
In this analysis, the polarized PDF's are optimized 
including precise SLAC-E155 proton target data \cite{E155p}.
Then, we examine an influence of the precise data 
on the polarized PDF uncertainty, 
which is estimated by the Hessian method.

\section{Parameterization of the polarized PDF's}
The polarized PDF's are determined by using spin asymmetry $A_1$ 
of the polarized DIS experiments 
from the EMC, SMC, SLAC-E130, E142, E143, E154, E155, and HERMES:
\begin{equation}
  A_1 (x, Q^2) = \frac{2x [ 1+R(x,Q^2) ] }{F_2(x,Q^2)} g_1(x,Q^2),
\end{equation}
where $F_2$ is the unpolarized structure function. 
The function $R(x,Q^2)=\sigma_L/\sigma_T$ is the ratio 
of absorption cross sections 
for longitudinal and transverse virtual photons, 
and it is determined from experimental data 
in reasonably wide $Q^2$ and $x$ ranges 
in the SLAC experiments \cite{R199X}. 
The polarized structure function $g_1$ is expressed with polarized PDF's:
\begin{equation}
  g_1(x,Q^2) = \frac{1}{2}\sum\limits_{i=1}^{n_f} e_{i}^2
     \bigg\{ \Delta C_q(x,\alpha_s) \otimes [ \Delta q_{i} (x,Q^2)
           + \Delta \bar{q}_{i} (x,Q^2) ] + \Delta C_g(x,\alpha_s)
    \otimes  \Delta g (x,Q^2) \bigg \},
\end{equation}
where $e_i$ is the electric charge of quarks,
and $\Delta C_q$, $\Delta C_g$ are Wilson's coefficient functions. 
The convolution $\otimes$ is defined by 
$ f (x) \otimes g (x) = \int^{1}_{x} dy/y\ f(x/y) g(y)$.
The polarized PDF's $\Delta f (\equiv f^{\uparrow}-f^{\downarrow} )$ 
are defined as helicity distributions in the nucleon.
In the AAC analysis, 
the polarized PDF $\Delta f(x)$ is defined at initial $Q^2$ 
by the weight function form:
\begin{equation}
	\Delta f(x)=A x^{\alpha}(1+\lambda x^{\gamma}) f(x) \ , 
\end{equation}
where $f(x)$ is the unpolarized PDF, 
and $A$, $\alpha$, $\lambda$, and $\gamma$ are free parameters.
Optimized PDF's are four distributions; 
$\Delta u_v(x)$, $\Delta d_v(x)$, $\Delta \bar{q}(x)$, and $\Delta g(x)$, 
and these are evolved from the initial $Q^2$(=1 GeV$^2$) 
to the same $Q^2$ of experimental data by the DGLAP equation \cite{DGLAP}.
In particular, the gluon distribution $\Delta g(x)$ contributes 
to the structure function 
with the non-zero coefficient function $\Delta C_g$ 
in the NLO case.

This analysis uses two constraint conditions.
First, the positivity condition is used to restrict large-$x$ behavior 
of the polarized PDF's.
This condition corresponds to the probabilistic interpretation 
of the parton distributions in the LO: $|\Delta f(x)| \le f(x)$.
It needs not to be satisfied strictly in the NLO analysis.
However, the polarized antiquark and gluon distributions tend 
to badly break the positivity limit: $|\Delta f(x)| \gg f(x)$.
Such excessive behavior is due to the large experimental errors 
in the large-$x$ region.
Hence, this behavior should be limited by this condition.

Next, the SU(3)$_f$ flavor symmetry is assumed:
$\Delta \bar{u}(x)=\Delta \bar{d}(x)=\Delta \bar{s}(x)=\Delta s(x)$.
Using this assumption, 
one can fix the first moments of the valence quarks 
with hyperon decay constants, 
then $\Delta u_v=0.926$ and $\Delta d_v=-0.341$ are obtained.
Note that the Bjorken sum rule is satisfied automatically 
by fixing first moments.
Furthermore, the spin content $\Delta \Sigma$ is obtained by 
$\Delta \Sigma_{N_f=3}=\Delta u_v+\Delta d_v+6\Delta \bar{q}$.
Since, the antiquark contribution is emphasized, 
then the spin content determination is susceptible to 
the antiquark behavior. 

In the analysis, 
we choose the modified minimal subtraction ($\overline{\rm MS}$) scheme, 
and the GRV parameterization for the unpolarized PDF's 
at the NLO analysis \cite{GRV98}.
The total $\chi^2$ is minimized by the CERN subroutine {\tt MINUIT}. 

\section{Uncertainty estimation}
Fortunately PDF uncertainty estimation method has been developed 
in the last several years 
(see a brief review \cite{Botje-E}).
The polarized PDF uncertainty comes from several error sources, 
e.g., experimental errors, unpolarized PDF, $\Lambda_{QCD}$, and so on.
However, it is difficult to incorporate these errors 
into uncertainty estimation simultaneously.
In the present analysis, 
the polarized PDF uncertainty is estimated from experimental errors 
by using the Hessian matrix $H_{ij}$ 
which is defined as a second order derivative matrix 
in the expanded $\chi^2(a_i)$ function around its minimum point.
The PDF uncertainty $\delta \Delta f(x)$ can be obtained easily 
by the inverse matrix of the Hessian and linear error propagation:
\begin{equation}
	[\delta \Delta f(x)]^2=\Delta \chi^2 \sum_{i, j}
	  \frac{\partial \Delta f(x)}{\partial a_i}
	  H_{ij}^{-1}
	  \frac{\partial \Delta f(x)}{\partial a_j}\ ,
\end{equation}
where $\Delta \chi^2(=\chi^2(a_i)-\chi^2_{min})$ is defined 
as the difference from the minimum $\chi^2$. 
It determines a confidence level of the PDF uncertainty, 
and it depends on the $\chi^2$ distribution $K(s)$ 
with \textit{N} degrees of freedom.
Here, \textit{N} is the number of optimized parameters.
In our estimation, 
the value of $\Delta \chi^2$ is obtained by the following equation:
$ \int^{\Delta \chi^2}_0 K(s) \ ds = \sigma \ , $
where $\sigma(=0.683)$ corresponds to 1 $\sigma$ error 
of a standard distribution 
in order to compare with general experimental errors.
The statistical and systematic errors are added in quadrature, 
so that it could be overestimation.
The proper estimation exists between the overestimated uncertainty 
and the uncertainty from only the statistical error.

\section{Results and discussions}
\begin{figure}[b!]
		\includegraphics*[width=60mm]{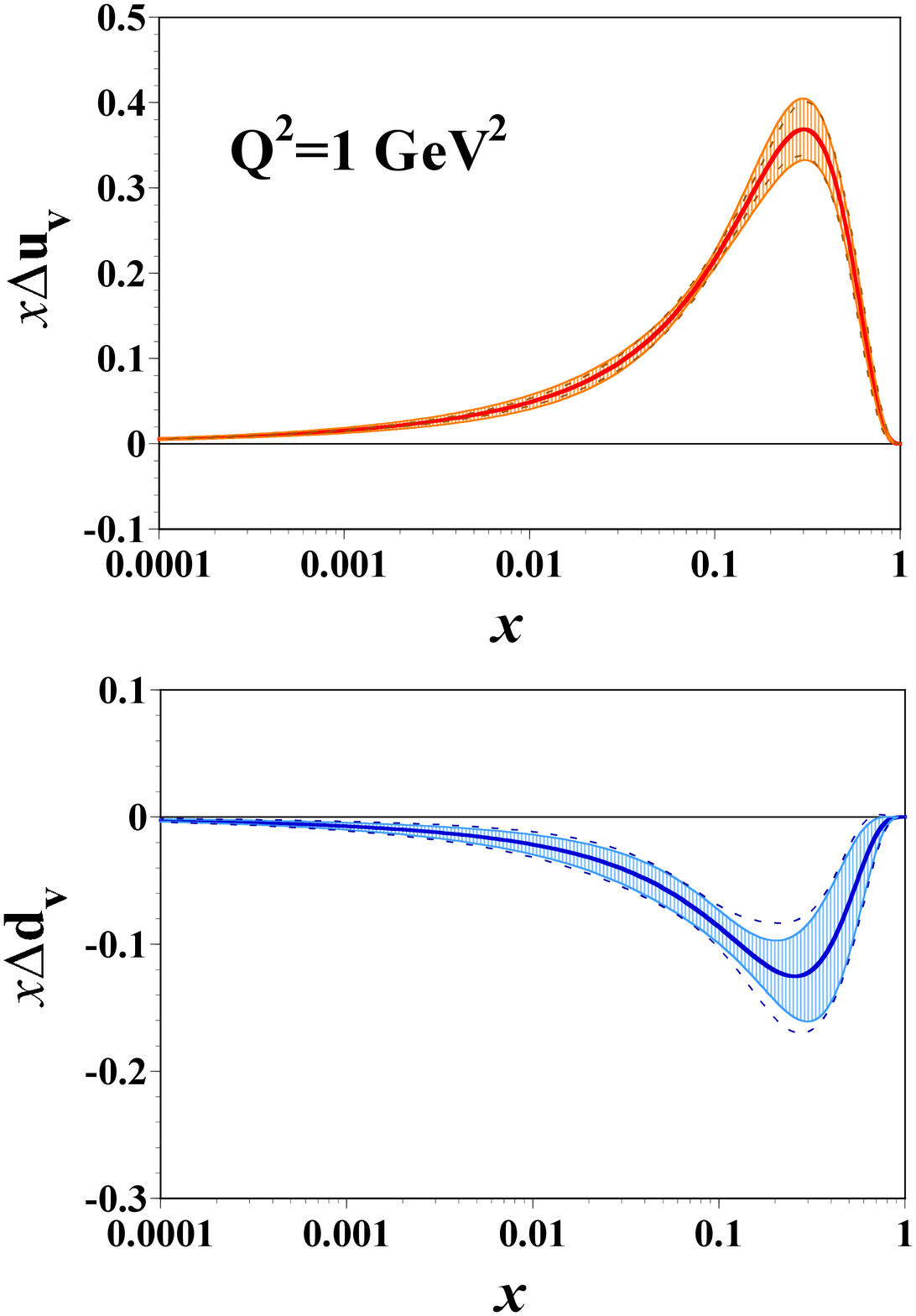}
	\hspace{6mm}
		\includegraphics*[width=60mm]{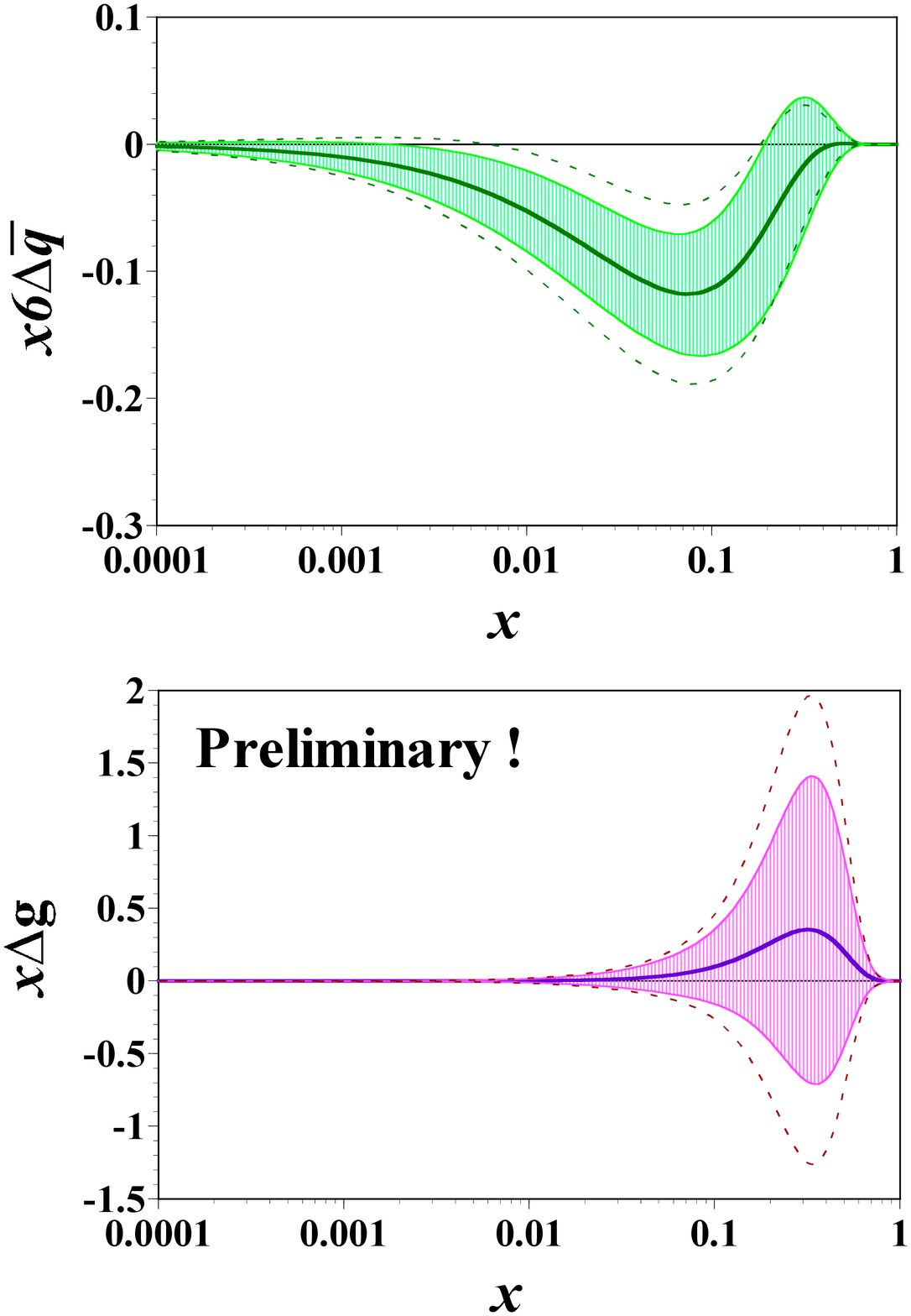} \\
\caption{Polarized PDF's with their uncertainties at $Q^2=1$ GeV$^2$. 
Dashed curves are the uncertainties of previous results (AAC NLO-2)}
	\label{fig:df-nlo1}
\end{figure}

The best fitting result is $\chi^2(/d.o.f.)=346.33(0.90)$.
The first moments of new results and the AAC pervious results 
(AAC00, NLO set2) \cite{AAC} 
are shown in Table. \ref{1stm}. 
A correlation coefficient $\rho_{\bar{q}g}$ 
between the first moment of the antiquark and gluon distributions
is $\rho_{\bar{q}g}=-0.836$, 
and there is strong correlation between two distributions.
The uncertainties of the new results become smaller than 
those of the previous results.
The gluon first moment and spin content $\Delta \Sigma$ 
still have large uncertainty.
The fixed first moments $\Delta u_v$ and $\Delta d_v$ 
do not have uncertainty, 
then the $\Delta \Sigma$ uncertainty is six times as large as 
the antiquark uncertainty. 
Thus, the spin content is subject to the uncertainty 
of the antiquark distribution.
Figure \ref{fig:df-nlo1} shows the uncertainty 
of the new antiquark distribution.
The antiquark uncertainty becomes rather large 
in the region $x< 0.01$, 
however the experimental data scarcely exist. 
The polarized DIS spin asymmetries $A_1^{p,d}(x)$ approaches rapidly 
to zero in the rang $x<0.004$.
It is insufficient to clarify small-$x$ behavior 
of the antiquark distribution.
Therefore, the antiquark determination has extrapolating ambiguity 
in small-$x$ region. 
It is needed tight constraint condition or other experiment.

In addition, Figure \ref{fig:df-nlo1} shows comparison between 
the PDF uncertainties of new results and the previous results.
There are no significant improvements of the valence quark uncertainties.
On the SU(3)$_f$ symmetry assumption, 
the fixing first moments strongly restricts the behavior
of valence quark distributions. 
In contrast, the antiquark and gluon uncertainties are reduced 
in the range $0.01<x<0.5$, where the E155 proton data exist.
The precise polarized DIS data can reduce the antiquark uncertainty mainly.
On the other hand, 
the gluon uncertainty changes in response to 
antiquark uncertainty reduction 
due to a strong correlation between two distributions.
Since the gluon contribution to the structure function $g_1(x)$ 
is smaller than the quark and antiquark contributions, 
we can extract only a little information of the gluon distribution 
in spite of the NLO analysis.
Actually, the gluon uncertainty is still large.
It indicates the difficulty of determining the gluon distributions 
from the polarized DIS data. 
Therefore, the uncertainty reduction of the gluon distribution is 
due to the strong correlation rather than the NLO contribution.
\begin{table}[t!]
\caption{First moments of the polarized antiquark, gluon, 
and spin content $\Delta \Sigma$ with their uncertainties at $Q^2=$1 GeV$^2$.}
\vspace{1mm}
\begin{tabular}{c|ccc}\hline
	& $\Delta \bar{q}$  & $\Delta g$         & $\Delta \Sigma$   \\
 \hline
New   & $-0.062 \pm 0.023$, & $0.499 \pm 1.268$, & $0.213 \pm 0.138$ \\ 
AAC00 & $-0.057 \pm 0.038$, & $0.532 \pm 1.949$, & $0.241 \pm 0.228$ \\
\hline
\end{tabular}	
\label{1stm}
\end{table}
\noindent

\begin{figure}[t!]
	\includegraphics*[width=60mm]{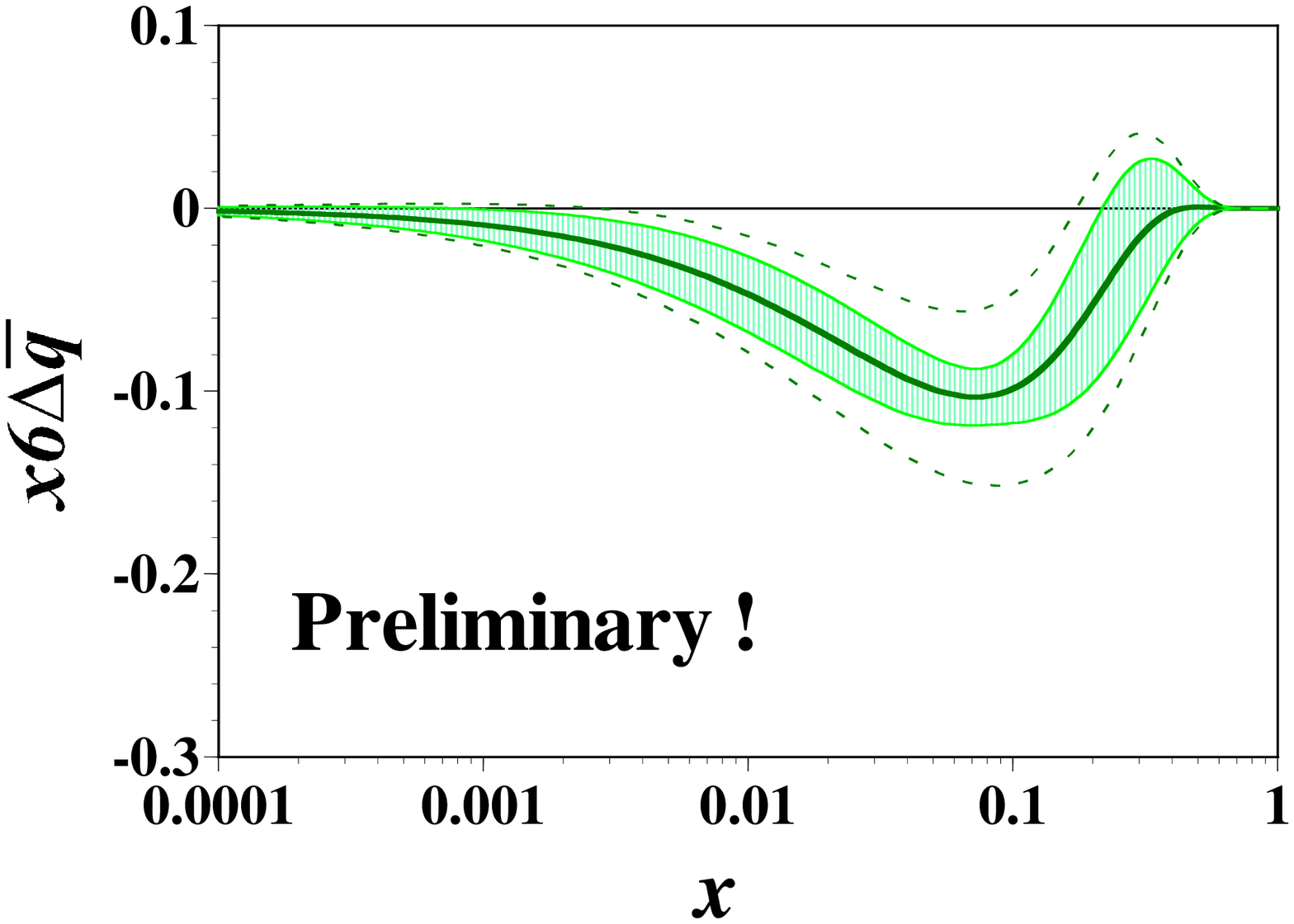}
	\hspace{6mm}
	\includegraphics*[width=60mm]{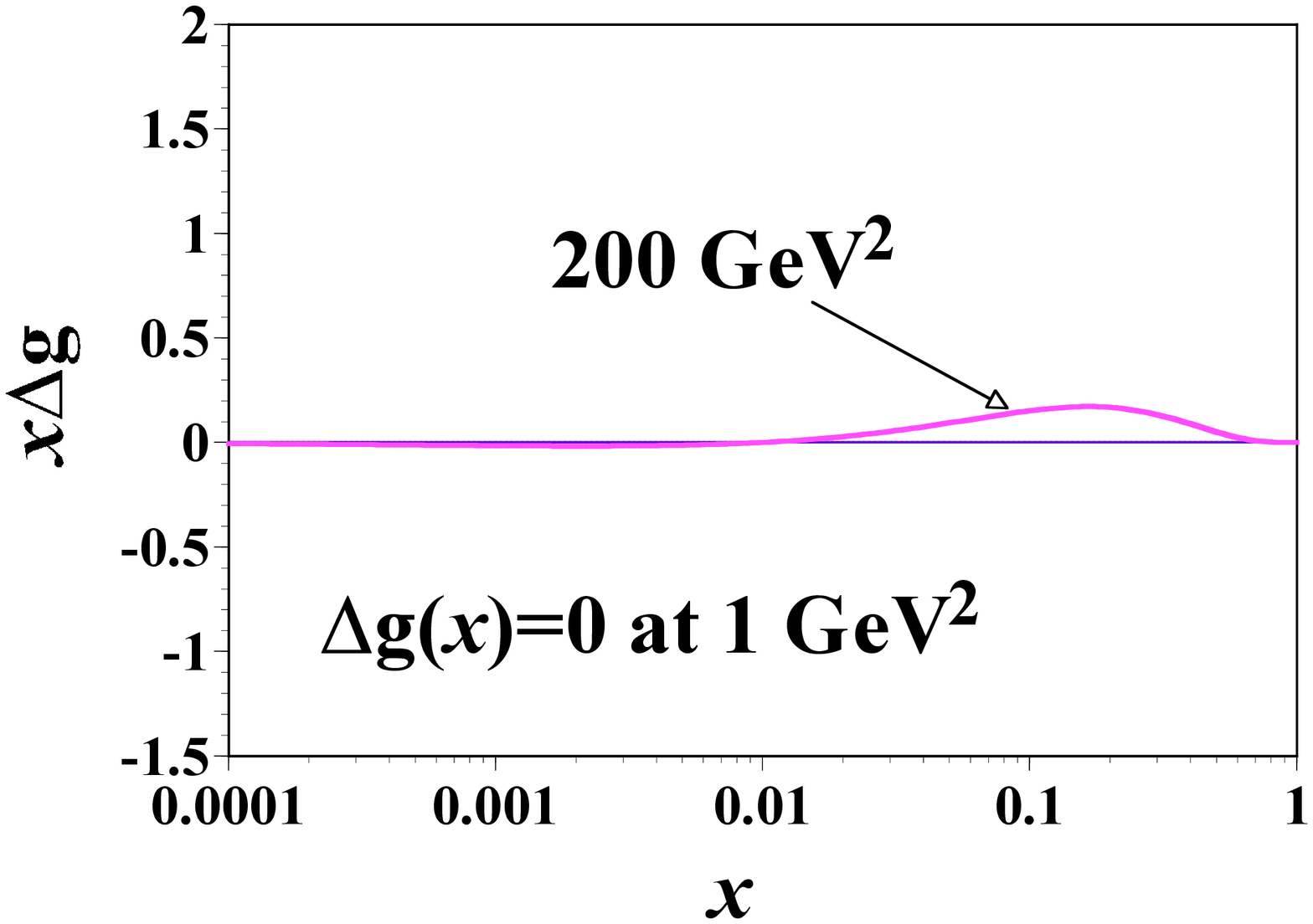} \\
\caption{Polarized antiquark and gluon distributions 
with their uncertainties at $Q^2=1$ GeV$^2$. 
The shaded portion shows the uncertainty of $\Delta g(x) = 0$ results, 
and the dashed curves are the uncertainties of new results ($\Delta g(x)\neq 0$).}
	\label{fig:dg0-nlo1}
\end{figure}

In order to examine the correlation effect on the parameterization,
we re-analyzed the $\Delta g(x)=0$ case 
in which the fixed gluon distribution does not have uncertainty.
The polarized PDF uncertainties of the $\Delta g(x)=0$ case are compared 
to those of the $\Delta g(x)\neq 0$ case in Figure \ref{fig:dg0-nlo1}.
The gluon distribution slightly exists at high-$Q^2$ 
due to $Q^2$ evolution of the singlet type DGLAP equation.
The valence quark uncertainties scarcely change.
Drastic improvement of the antiquark uncertainty is due to 
vanished the large gluon uncertainty.
the obscure gluon distribution brings about 
the larger antiquark uncertainty by the complementary relation. 

\section{Summary}
By this analysis, the polarized PDF's were optimized 
from the polarized DIS world data 
which included the SLAC-E155 proton target data. 
The polarized PDF uncertainties were estimated by the Hessian method.
The E155 precise measurements scarcely improve the valence quark uncertainties,
but they can reduce the antiquark and gluon uncertainties.
These, however, are still wrapped in large uncertainty.
The SU(3)$_f$ symmetry, which we are obliged to assume, 
restricts strongly the valence quark behavior by fixing first moments, 
and the spin content determination depends on the antiquark behavior.
Additionally, there is the strong correlation 
between the antiquark and gluon distributions.
If the gluon distribution is clarified by RHIC-Spin at BNL, 
the uncertainty of the antiquark distribution can be reduced to some extent.
Similarly, the complementary relation can reduce the
large uncertainty of the spin content 
which comes from the extrapolating issue of the antiquark behavior.
Then, we will be able to investigate the antiquark flavor dependence 
in detail.


\begin{thebibliography}{00}
\bibitem{AAC}
	AAC, Y. Goto {\it et al}., Phys. Rev. {\bf D62} (2000) 034017. 

\bibitem{polpdf}
	De Florian and R. Sassot, Phys. Rev. D62 (2000) 094025; 
	M. Gl\"uck, E. Reya, M. Stratmann, and W. Vogelsang, 
		Phsy. Rev. {\bf D63} (2001) 09400; 
	E. Leader, A.V. Sidorov, and D.B. Stamenov, 
		Eur.Phys.J. C23 (2002) 479-485; 
	J. Bl\"umlein and H.  B\"ottcher, 
		Nucl. Phys. B {\bf B636} (2002) 225-263. 
	Fortran program librarys of polarized PDF's are available from 
	http://www-spires.dur.ac.uk/hepdata/pdf.html.

\bibitem{SDIS} 
	SMC, B. Adeva {\it et al}., 
		Phys. Lett. {\bf B420}, 180 (1998);
	HERMES, K. Ackerstaff {\it et al}., 
		Phys. Lett. {\bf B464} 123 (1999).

\bibitem{E155p}
	SLAC-E155, P. L. Anthony {\it et al}., 
		Phys. Lett. {\bf B493} (2000) 19.

\bibitem{R199X}
	L. W. Whitlow, S. Rock, A. Bodek, S. Dasu and E. M. Riordan, 
		Phys. Lett. {\bf B250}, 193 (1990);
	SLAC-E143, K. Abe {\it et al}., Nucl. Phys {\bf B452} (1999) 194

\bibitem{Botje-E}
	M. Botje , J. Phys. {\bf G} 28 (2002) 779-790.

\bibitem{DGLAP} 
	V. N. Gribov and L. N. Lipatov, 
		Sov. J. Nucl. Phys. 15 (1972) 438 and 675;
	G. Altarelli and G. Parisi, Nucl. Phys. B 126 (1977) 298;
	Yu. L. Dokshitzer, Sov. Phys. JETP 46 (1977) 641.

\bibitem{GRV98}
	M. Gl\"uck, E. Reya, and A. Vogt, 
		Eur. Phys. J. {\bf C5} (1998) 461. 
\end{thebibliography}
\end{document}